\documentclass[11pt, a4paper]{article}

\usepackage{jheppub}
\usepackage{amsmath}
\usepackage{amsfonts}
\usepackage{amssymb} 
\usepackage{mathrsfs}
\usepackage[utf8]{inputenc}
\usepackage{hyperref}
\usepackage{mathtools}

\title{Super Yang-Mills Action from Hybrid Superstring Field Theory}
\author{Ulisses M. Portugal}

\affiliation{ICTP South American Institute for Fundamental Research\\Instituto de F\'{i}sica Te\'{o}rica, UNESP — Universidade Estadual Paulista,\\Rua Dr. Bento T. Ferraz 271, 01140-070, S\~{a}o Paulo, SP, Brasil}
\emailAdd{ulisses.portugal@unesp.br}

\abstract{We explicitly compute the effective action from Open Superstring Field Theory in the hybrid formalism to quartic order in the $\alpha'\rightarrow 0$ limit, and show that it reproduces ten-dimensional Super Yang-Mills in terms of four-dimensional superfields. We also show that in this limit the gauge transformations coincide with SYM to all orders, which means that the effective action should reproduce SYM to all orders.}

\begin{document}

\maketitle

\section{Introduction}
String Field Theory provides an off-shell description of string theory. As such, it's expected that, upon integrating out the massive fields, we should obtain an effective field theory describing the low-energy limit of string theory. Effective actions were computed from bosonic open string field theory in \cite{Taylor:2000ek,Coletti:2003ai} using level truncation. This was generalized to Open Superstring Field Theory (OSFT) in \cite{Berkovits:2003ny}, where the authors found a method that allows one to integrate out all massive states at once, no truncation required. Starting from the WZW-like formulation for the NS sector of OSFT \cite{Berkovits:1995ab}, they computed the effective action to quartic order, and showed that it reproduces the Yang-Mills action, up to $\alpha'$ corrections (not super Yang-Mills because the R sector was not included). One can argue by dimensional analysis that there cannot be any higher order terms at $\alpha'\rightarrow 0$, so that the effective theory coincides exactly with Yang-Mills in this limit.

There is an alternative formulation of OSFT using the hybrid formalism \cite{Berkovits:1996bf,Berkovits:1995ab}, which is convenient for describing strings compactified on a Calabi-Yau manifold. In addition to having manifest four-dimensional supersymmetry, this formulation has the advantage of automatically including the R sector. The effective theory resulting from this formulation is expected to reproduce ten-dimensional super Yang-Mills (SYM) in terms of four-dimensional superfields \cite{Marcus:1983wb} (plus $\alpha'$ corrections). In this paper we will use the methods of \cite{Berkovits:2003ny} to check that this is true at the quartic level. Although we do not compute the full action explicitly, we will argue from gauge invariance that it must reproduce SYM.

In section \ref{sec Hybrid formulation of OSFT} we will review the essential definitions of the hybrid formalism and the OSFT action. In section \ref{sec Massless contribution to the effective action} we compute the massless contribution to the effective action, and find that it coincides with the SYM action to cubic order, but not quartic. In section \ref{sec Massive contribution} we integrate out the massive levels exactly, and find that, upon adding this contribution, the effective action coincides with SYM up to quartic order. In section \ref{sec Gauge invariance and the full action} we obtain the full gauge transformations at $\alpha'\rightarrow 0$ and argue that it implies that the full effective action is the same as SYM in this limit.

\section{Hybrid formulation of OSFT}\label{sec Hybrid formulation of OSFT}
In this section we will briefly present the relevant definitions and results of the hybrid formalism. For details, see \cite{Berkovits:1996bf,Berkovits:1995ab}. The hybrid variables consist of eleven free bosons ($x^m,\rho,x^j,\bar x_j$) and fourteen free fermions $(\theta^{\alpha}, \bar{\theta}^{\dot{\alpha}},$
$ p^{\alpha}, \bar{p}^{\dot{\alpha}}, \psi^j,\bar\psi_j)$ with OPE's
\begin{gather}
    x^{m}(z_1) x^{n}(z_2) \sim -\log |z_1-z_2| \delta^{m n}, \quad \rho(z_1) \rho(z_2) \sim -\log (z_1-z_2) \\
    p_{\alpha}(z_1) \theta^{\beta}(z_2) \sim \delta_{\alpha}^{\beta}(z_1-z_2)^{-1}, \quad \bar{p}_{\dot{\alpha}}(z_1) \theta^{\dot{\beta}}(z_2) \sim \delta_{\dot{\alpha}}^{\dot{\beta}}(z_1-z_2)^{-1} \nonumber \\
    x^{i}(z_1) \bar x_{j}(z_2) \sim -\log |z_1-z_2|\delta^i_j, \quad \psi^{i}(z_1) \bar \psi_{j}(z_2) \sim (z_1-z_2)^{-1}\delta^i_j \nonumber
\end{gather}
Here $m,\alpha,\dot\alpha$ are four dimensional vector and spinor indices, while $i,j=1$to $3$ are $SU(3)$ indices corresponding to a six-dimensional compactified manifold.

We define a set of twisted $\mathrm{N}=4$ superconformal generators in terms of these variables as
\begin{gather}
    T=\frac{1}{2} \partial x^{m} \partial x_{m}+p_{\alpha} \partial \theta^{\alpha}+\bar{p}_{\dot{\alpha}} \partial \bar{\theta}^{\dot{\alpha}}+\frac{1}{2} \partial \rho \partial \rho-\frac{i}{2} \partial^{2} \rho +\partial x^j \partial \bar{x}_j+i \psi^j \partial \bar{\psi}_j \\
    G^{+}=e^{\rho} d^2+\psi^j \partial \bar{x}_j, \quad G^{-}=e^{-\rho} \bar{d}^2+\bar{\psi}_j \partial x^j \\
    \tilde{G}^{+}=e^{-2 \rho+i H_{C}} \bar{d}^2+\frac{1}{2} e^{-\rho} \epsilon_{j k l} \psi^j \psi^k \partial x^l, \quad
    \tilde{G}^{-}=e^{2 \rho-i H_{C}} d^2+\frac{1}{2} e^{\rho} \epsilon^{j k l} \bar\psi_j \bar\psi_k \partial\bar x_l \\
    J=-\partial \rho+iJ_C, \quad J^{++}=e^{-\rho+i H_{C}}, \quad J^{--}=e^{\rho-i H_{C}}
\end{gather}
where
\begin{gather}
    d_{\alpha}=p_{\alpha}+\frac{i}{2} \bar{\theta}^{\dot{\alpha}} \partial x_{\alpha \dot{\alpha}}-\frac{1}{4}(\bar{\theta})^{2} \partial \theta_{\alpha}+\frac{1}{8} \theta_{\alpha} \partial(\bar{\theta})^{2} \\
    \bar{d}_{\dot{\alpha}}=\bar{p}_{\dot{\alpha}}+\frac{i}{2} \theta^{\alpha} \partial x_{\alpha \dot{\alpha}}-\frac{1}{4}(\theta)^{2} \partial \bar{\theta}_{\dot{\alpha}}+\frac{1}{8} \bar{\theta}_{\dot{\alpha}} \partial(\theta)^{2}
\end{gather}
$J_{C}=\psi^j \bar{\psi}_j=\partial H_{C}$ and $x_{\alpha \dot{\alpha}} = \sigma_{\alpha \dot{\alpha}}^{m} x_{m}$, $\sigma_{\alpha \dot{\alpha}}^{m}$ being the Pauli matrices. These can be related to RNS variables by a field redefinition, and then $G^+$ corresponds to the BRST current, $\tilde G^+$ to $\eta$ and $G^-$ to $b$. Hermitian conjugation is defined in such a way that $G^{\pm\dagger} = \tilde G^{\pm}$.

It's useful to break these generators into ``four-dimensional" and ``six-dimensional" parts as, for example
\begin{gather}
    G^{+}_4=e^{\rho} d^2, \quad G_6^+ = \psi^j \partial \bar{x}_j, \quad
    \tilde{G}^{+}_4 = e^{-2 \rho+i H_{C}} \bar{d}^2, \quad \tilde G_6^+ = \frac{1}{2} e^{-\rho} \epsilon_{j k l} \psi^j \psi^k \partial x^l \\
    G^{-}_4=e^{-\rho} \bar d^2, \quad G_6^- = \bar\psi_j \partial x^j, \quad
    \tilde{G}^{-}_4 = e^{2 \rho-i H_{C}} d^2, \quad \tilde G_6^- = \frac{1}{2} e^{\rho} \epsilon^{j k l} \bar\psi_j \bar\psi_k \partial\bar x_l
\end{gather}

The basic correlators are given by
\begin{align}
    \left\langle\theta^{2}\bar{\theta}^{2}\right\rangle_\theta &= 1 \\
    \left\langle e^{-\rho+i H_{C}}\right\rangle_{\rho,\psi} &= 1 \label{basic correlator} \\
    \left\langle e^{i k_{1} x}\left(z_{1}\right) e^{i k_{2} x}\left(z_{2}\right)\right\rangle_x &= i(2 \pi)^{4} \delta^{4}\left(k_{1}+k_{2}\right)\left|z_{1}-z_{2}\right|^{-k_{1} \cdot k_{2}}
\end{align}

\subsection{Hybrid open superstring field action}
The hybrid formulation of OSFT involves three string fields, separated by their $\rho$-charges - i.e. their eigenvalues under $-\int \partial \rho$. We denote them by $\Phi_{-1}$, $V\equiv\Phi_{0}$ and $\Phi_{1}$, where the subscript denotes the $\rho$-charge. They all have ghost number ($J$-charge) zero. The OSFT action is given by \cite{Berkovits:1995ab}:
\begin{align}\label{action}
    S_{Hybrid} = \frac{1}{2} \left<\left(e^{-V} G_4^{+} e^V\right)\left(e^{-V} \tilde{G}_4^{+} e^V\right)-\int_0^1 d t\left(e^{-t V} \partial_t e^{t V}\right)\left\{e^{-t V} G_4^{+} e^{t V}, e^{-t V} \tilde{G}_4^{+} e^{t V}\right\}\right. \nonumber\\
    +\left(e^{-V} G_6^{+} e^V\right)\left(e^{-V} \tilde{G}_6^{+} e^V\right)-\int_0^1 d t\left(e^{-t V} \partial_t e^{t V}\right)\left\{e^{-t V} G_6^{+} e^{t V}, e^{-t V} \tilde{G}_6^{+} e^{t V}\right\} \nonumber\\
    +2\left(\left(\tilde{G}_6^{+} e^{-V}\right) \bar{\Omega} e^V+e^V \Omega\left(G_6^{+} e^{-V}\right)+e^{-V} \bar{\Omega} e^V \Omega\right) \nonumber\\
    \left.-\left(\Omega \tilde{G}_6^{+} \Phi_1-\frac{2}{3} \Omega \Omega \Phi_1\right)+\left(\bar{\Omega} G_6^{+} \Phi_{-1}+\frac{2}{3} \bar{\Omega} \bar{\Omega} \Phi_{-1}\right)\right> .
\end{align}
where $\Omega=\tilde G_4^+\Phi_1$ and $\bar\Omega=G_4^+\Phi_{-1}$. To simplify the notation, we use the same symbol for the superconformal generators and for their zero modes when acting on the string fields - for example, $G_4^+ V$ denotes the zero mode of $G_4^+$ acting on $V$. The fields are multiplied with Witten's star product.

We expect that the effective action obtained upon integrating out all the massive fields should correspond to the action for 10-dimensional super-Yang-Mills in terms of four-dimensional superfields \cite{Marcus:1983wb}
\begin{gather}\label{MSS action}
    S_{SYM} = \text{Tr}\frac{1}{2} \int d^{10} x\left[-2 \int d^2 \theta W^\alpha W_\alpha\right. \nonumber\\
    \left.+\int d^4 \theta\left(\left(e^{-v} \bar{\partial}_j e^v\right)\left(e^{-v} \partial^j e^v\right)-\int_0^1 d t\left(e^{-t v} \partial_t e^{t v}\right)\left\{e^{-t v} \bar{\partial}_j e^{t v}, e^{-t v} \partial^j e^{t v}\right\}\right)\right) \nonumber\\
    +2 \int d^4 \theta\left(\left(\partial^j e^{-v}\right) \bar{\omega}_j e^v+e^v \omega^j\left(\bar{\partial}_j e^{-v}\right)+e^{-v} \bar{\omega}_j e^v \omega^j\right) \nonumber\\
    \left.+\int d^2 \theta \epsilon_{j k l}\left(\omega^j \partial^k \omega^l+\frac{2}{3} \omega^j \omega^k \omega^l\right)+\int d^2 \bar{\theta} \epsilon^{j k l}\left(\bar{\omega}_j \bar{\partial}_k \bar{\omega}_l-\frac{2}{3} \bar{\omega}_j \bar{\omega}_k \bar{\omega}_l\right)\right] .
\end{gather}
where $W_{\alpha} = \bar D^2(e^{-v}D_{\alpha}e^v)$. $v$ is the real superfield containing the four-dimensional part of the gauge field, and $\omega_j$ and $\bar{\omega}^j$ are the chiral and anti-chiral superfields containing the six-dimensional part of the gauge field. Note that all fields are allowed to depend on the full ten dimensions.

\section{Massless contribution to the effective action}\label{sec Massless contribution to the effective action}
Now we start computing the low-energy effective action. We will supress $\alpha'$ corrections. This means we only consider terms with two derivatives (where $D$ and $\bar D$ count as ``half a derivative"). In this section we compute the massless contribution up to quartic order - i.e. we simply truncate the string fields and their products to massless order in \ref{action}. This should result in the SYM action up to cubic order, but there should be a mismatch at quartic order, since at this order we will also have propagation of massive fields. We will now show that this is indeed the case.

At the massless level, the three string fields take the forms
\begin{equation}
    V=V(x,\theta,\bar\theta), \quad \Phi_{-1} = e^{-\rho}\psi^j\bar\sigma_j(x,\theta,\bar\theta), \quad \Phi_{1} = e^{\rho}\bar\psi_j\sigma^j(x,\theta,\bar\theta)
\end{equation}
where $x$ stands for all the ten directions. We expect that $V$ should be the same as the $v$ in \ref{MSS action}, while $\Phi_1$ and $\Phi_{-1}$ should be related to the chiral and anti-chiral superfields, in a manner to be made precise shortly. Acting with the $G$'s, we get
\begin{gather}
    G_4^+V = 2e^{\rho}dDV + \partial(e^{\rho})D^2V - ie^{\rho}\partial\bar\theta_{\dot\alpha}\partial^{\dot\alpha\alpha}D_{\alpha}V, \label{G4}\\
    \tilde G_4^+V = 2e^{-2\rho+iH}\bar d\bar DV + \partial(e^{-2\rho+iH})\bar D^2V - ie^{-2\rho+iH}\partial\theta_{\alpha}\partial^{\dot\alpha\alpha}\bar D_{\dot\alpha}V \label{tilde G4}\\
    G_6^+ V = \psi^j\partial_jV, \quad \tilde G_6^+ V = e^{-\rho}\frac{1}{2}\epsilon_{ijk}\psi^i\psi^j\bar\partial^kV \\
    \Omega = \frac{1}{2}e^{-\rho}\epsilon_{ijk}\psi^i\psi^j\bar D^2\sigma^k, \quad \bar\Omega = \psi^jD^2\bar\sigma_j \\
    \tilde G_6^+\Phi_1 = \epsilon_{ijk}\psi^i\bar\partial^j\sigma^k, \quad G_6^+\Phi_{-1} = e^{-\rho}\psi^i\psi^j\partial_i\bar\sigma_j \label{G_6}
\end{gather}
Now we see that $\Omega$ and $\bar\Omega$ contain a chiral and an anti-chiral superfield, respectively. Presumably $\bar D^2\sigma^i = \omega^i$ and $D^2\bar\sigma_i = \bar\omega_i$.

We start by computing the third term in \ref{action} for massless V. From the relations above, we have
\begin{equation}
    \left(e^{-V} G_6^{+} e^V\right)\left(e^{-V} \tilde{G}_6^{+} e^V\right) = \left(e^{-V} \bar{\partial}^i e^V\right)\left(e^{-V} \partial_l e^V\right)\frac{1}{2}e^{-\rho}\epsilon_{ijk}\psi^j\psi^k\psi^l
\end{equation}
The $e^{-\rho}\psi^3$ factor gives precisely the basic correlator \ref{basic correlator}, and all the OPE's are regular. Therefore, we obtain precisely the corresponding term in the super-Yang-Mills action. The same logic follows for all terms in \ref{action} but the first two. These are a bit more involved, since we have to take the OPE's of the $d$'s and $\bar d$'s coming from $G_4^+$ and $\tilde G_4^+$, as well as the $\rho$ exponentials. Thus we see that all the terms in \ref{MSS action} involving the chiral or anti-chiral superfields or the six-dimensional derivatives of $v$ are matched exactly by the truncation of \ref{action}, to all orders. Next we will investigate the first two terms in \ref{action}.

\subsection{Cubic term}
The cubic term in the four dimensional part of the SFT action is
\begin{equation}\label{SFT cubic}
    S^{(3)} = -\frac{1}{6}\left< V\{G_4^+V,\tilde G_4^+V\}\right>
\end{equation}
which is to be compared with the SYM cubic term
\begin{equation}\label{SYM cubic}
    S^{(3)}_{SYM} = -\text{Tr}\int d^{10}xd^2\theta\,\bar D^2[D^{\alpha}V,V]\bar D^2D_{\alpha}V = -\text{Tr}\int d^{10}xd^4\theta\,[D^{\alpha}V,V]\bar D^2D_{\alpha}V
\end{equation}
where we have transformed a total $\bar D^2$ derivative into a $d^2\bar\theta$ integral. This integral is antysimmetric with respect to flipping the chiralities (i.e., $D\leftrightarrow \bar D$), and it will be convenient to write it in a way that makes this manifest. Using integration by parts and cyclicity, one can show that
\begin{equation}\label{SYM cubic 2}
    S^{(3)}_{SYM} = \text{Tr}\int d^{10}xd^4\theta\, V[\bar D^2V,D^2V] + 2V[D^{\alpha}\bar D_{\dot\alpha}V,\bar D^{\dot\alpha}D_{\alpha}V]
\end{equation}

Now we compute \ref{SFT cubic} for massless V. We will compute the correlators in the upper half plane (UHP), where the prescription for the star product is that the operators be inserted at $-\sqrt{3}$, $0$ and $\sqrt{3}$. Using \ref{G4}, \ref{tilde G4} we get
\begin{align}\label{1st cubic term}
    \left<G_4^+VV\tilde G_4^+V\right> = 48\left<dDVV\bar d\bar DV\right> - 8\sqrt{3}(\left<D^2VV\bar d\bar DV\right>-\left<dDVV\bar D^2V\right>) \\ 
    - 2\left<D^2VV\bar D^2V\right> - 24i(\left<dDVV\partial\theta_{\alpha}\partial^{\dot\alpha\alpha}\bar D_{\dot\alpha}V\right> + \left<\partial\bar\theta_{\dot\alpha}\partial^{\dot\alpha\alpha}D_{\alpha}VV\bar d\bar DV\right>) \nonumber
\end{align}
Here we hid the positions of the operators for ease of notation. On the r.h.s., we already computed the $\rho$ and $\psi$ correlators. Wherever there is no $\rho$ or $\psi$ dependence, the brackets are to be understood as just the $x$ and $\theta$ correlators. The other term in \ref{SFT cubic} is given by hermitian conjugation, which translates into $D\leftrightarrow\bar D$ with a minus sign.

The correlators give
\begin{gather}
    \left<dDV(-\sqrt{3})V(0)\bar d\bar DV(\sqrt{3})\right> = \left<\frac{1}{6}\bar D^{\dot\alpha}D_{\alpha}VD^{\alpha}V\bar D_{\dot\alpha}V - \frac{1}{6}D_{\alpha}V\bar D^{\dot\alpha}VD^{\alpha}\bar D_{\dot\alpha}V \right. \\
     + \frac{1}{12}\bar D^{\dot\alpha}D_{\alpha}VVD^{\alpha}\bar D_{\dot\alpha}V + \frac{1}{6}D_{\alpha}V[D^{\alpha},\bar D^{\dot\alpha}]V\bar D_{\dot\alpha}V +\frac{i}{24}\partial_{\alpha\dot\alpha}D^{\alpha}VV\bar D^{\dot\alpha}V \nonumber\\
    \left. -\frac{i}{24}D^{\alpha}VV\partial_{\alpha\dot\alpha}\bar D^{\dot\alpha}V\right> \nonumber\\
    \left<dDV(-\sqrt{3})V(0)\bar D^2V(\sqrt{3})\right> = \frac{1}{\sqrt{3}}\left< D_{\alpha}VD^{\alpha}V\bar D^2V + \frac{1}{2}D_{\alpha}VVD^{\alpha}\bar D^2V\right> \\
    \left<dDV(-\sqrt{3})V(0)\partial\theta_{\alpha}\partial^{\dot\alpha\alpha}\bar D_{\dot\alpha}V(\sqrt{3})\right> = -\frac{i}{12}\left<D_{\alpha}VV\partial^{\dot\alpha\alpha}\bar D_{\dot\alpha}V\right>
\end{gather}
Now, we again use integration by parts and cyclicity to write this in a form in which there is either two or zero $D$'s or $\bar D$'s on each V, as in \ref{SYM cubic 2}.
Substituting in \ref{1st cubic term}, we obtain
\begin{equation}
    \left<G_4^+VV\tilde G_4^+V\right> = 12\left<\bar D^{\dot\alpha}D_{\alpha}VVD^{\alpha}\bar D_{\dot\alpha}V\right> + 6\left<D^2VV\bar D^2V\right>
\end{equation}
Adding the conjugate and the factor of $1/6$, this gives precisely \ref{SYM cubic 2}.

\subsection{Quartic term}
The SYM quartic term is symmetric with respect to $D\leftrightarrow\bar D$. Following the method used for the cubic term, one can show that it can be written as
\begin{gather}\label{SYM quartic}
    S^{(4)}_{SYM} = -\text{Tr}\frac{1}{12}\int d^{10}xd^4\theta\, [D^{\alpha},\bar D^{\dot\alpha}]V(\{D_{\alpha}V,[V,\bar D_{\dot\alpha}V]\} - \{\bar D_{\dot\alpha}V,[V,D_{\alpha}V]\}) \\
    - 2[D_{\alpha}V,\{\bar D^{\dot\alpha}V,D^{\alpha}V\}]\bar D_{\dot\alpha}V - [V,[V,D^2V]]\bar D^2V \nonumber\\
    + i([V,[V,D^{\alpha}V]]\partial_{\alpha\dot\alpha}\bar D^{\dot\alpha}V + [V,[V,\bar D^{\dot\alpha}V]]\partial_{\alpha\dot\alpha}D^{\alpha}V) \nonumber
\end{gather}

On the OSFT side, the quartic term for the four-dimensional part of the action is
\begin{equation}
    S_4^{(4)} = \frac{1}{4!}\left<G_4V\tilde G_4VV^2+G_4VV^2\tilde G_4V -2G_4VV\tilde G_4VV\right>
\end{equation}
One might think that computing this term would require a lot more work than the cubic one, but we will now argue that all the relevant quartic correlators can be immediately read off from the cubic ones, up to simple factors. First, note that now the operators should be inserted at $-1$, $0$, $1$ and $\infty$. By cyclicity, we can always choose to put a V with no G's at infinity, and then all contractions with this V vanish - it becomes just a spectator. Thus the correlators take the same form as the cubic ones, with an extra V inserted at infinity, and with a rescaling by $1/\sqrt{3}$. The result, after organizing the terms to look as close to \ref{SYM quartic} as possible, is
\begin{equation}\label{S4 massless}
    S^{(4)}_{Massless} = S^{(4)}_{SYM} + \frac{1}{4}\text{Tr}\int d^{10}xd^4\theta\,  D_{\alpha}V\bar D_{\dot\alpha}VD^{\alpha}V\bar D^{\dot\alpha}V - 2D_{\alpha}V\bar D_{\dot\alpha}V\bar D^{\dot\alpha}VD^{\alpha}V
\end{equation}
As expected, there is a mismatch. We will see in the next section that the extra terms are precisely cancelled by the contribution to the action from integrating out the higher mass levels.

\section{Massive contribution}\label{sec Massive contribution}
To integrate out the massive fields, we will use the methods of \cite{Berkovits:2003ny}, which allow us to obtain the exact answer, without any level truncation. We write the string fields as
\begin{equation}\label{decomposition}
    \Phi_i = \Phi_i^0 + R_i
\end{equation}
where $\Phi_i^0$ contain the massless states, and $R_i$ contain the fields we want to integrate out. We now solve the equations of motion for $R_i$ in terms of $\Phi_i^0$. Since we are only computing the quartic term in the action, we only need the solution to quadratic order. The equations of motion are \cite{Berkovits:1995ab}:
\begin{gather}
    (\tilde G_4^+ G_4^+ + \tilde G_6^+ G_6^+)V + \tilde G_6^+\bar\Omega - G_6^+\Omega - \frac{1}{2}\{\tilde G_4^+V, G_4^+V\} - \frac{1}{2}\{\tilde G_6^+V, G_6^+V\} \\
    + \frac{1}{2}[\tilde G_6^+\bar\Omega,V] + \frac{1}{2}[G_6^+\Omega,V] - \{\bar\Omega,\tilde G_6^+V\} - \{\Omega, G_6^+V\} = O(\Phi_i^3) \nonumber \\
    \tilde G_6^+ G_4^+V + G_6^+\bar\Omega - G_4^+\Omega - \frac{1}{2}\{\tilde G_6^+V, G_4^+V\}
    + \frac{1}{2}[G_6^+\bar\Omega,V] + \frac{1}{2}[G_4^+\Omega,V] \\
    - \{\Omega, G_4^+V\} + \bar\Omega^2 = O(\Phi_i^3) \nonumber \\
    \tilde G_4^+ G_6^+V + \tilde G_4^+\bar\Omega - \tilde G_6^+\Omega - \frac{1}{2}\{\tilde G_4^+V, G_6^+V\} + \frac{1}{2}[\tilde G_4^+\bar\Omega,V] + \frac{1}{2}[\tilde G_6^+\Omega,V] \\
    - \{\bar\Omega, \tilde G_4^+V\} + \Omega^2 = O(\Phi_i^3) \nonumber
\end{gather}
which are solved to quadratic order by
\begin{gather}\label{solution}
    R_{-1} = - \frac{1}{2}(\frac{G_4^-}{L_0}[V^0,G_6^+V^0]+\frac{G_4^-}{L_0}[V^0,G_4^+\Phi_{-1}^0]+\frac{G_6^-}{L_0}[\Phi_{-1}^0,G_4^+\Phi_{-1}^0]) \\
    R_{0} = - \frac{1}{2}(\frac{G_4^-}{L_0}[V^0,G_4^+V^0]+\frac{G_6^-}{L_0}[V^0,G_6^+V^0]+\frac{G_4^-}{L_0}[V^0,G_6^+\Phi_{1}^0]+\frac{G_6^-}{L_0}[V^0,G_4^+\Phi_{-1}^0]) \\
    R_{1} = - \frac{1}{2}(\frac{G_6^-}{L_0}[V^0,G_4^+V^0]+\frac{G_4^-}{L_0}[V^0,G_4^+\Phi_{1}^0]+\frac{G_4^-}{L_0}[\Phi_{1}^0,G_6^+\Phi_{1}^0])
\end{gather}
Now we obtain the contribution from $R_{i}$ to the quartic term by simply substituting these solutions into the $R_{i}$ kinetic terms. Note that we have now four $G$ operators, which will become simple derivatives on the massless fields, as in \ref{G4}-\ref{G_6}. Since we do not want the higher derivative corrections, we consider only the terms with two derivatives (again, $D$ and $\bar D$ count as half). That means that each $G$ operator can only contribute one $D$ or $\bar D$, which implies that the only nonzero contribution is from the four-dimensional term, namely
\begin{equation}\label{BPZ product}
    \left<[\tilde G_4^+V^0,G_4^+V^0],\frac{G_4^-}{L_0}[V^0,G_4^+V^0]\right>
\end{equation}
where $<,>$ denotes the BPZ product. This is as should be, since, as we have seen, the four-dimensional term is the only one for which the massless contribution does not match exactly the SYM action. Since the $R_i$'s will no longer appear, we will drop the $0$ superscript on $V$.

Now we use the crucial trick from \cite{Berkovits:2003ny}, which allows us to rewrite the BPZ product of two star products, such as \ref{BPZ product}, as a correlator with insertion points $-1/a$, $1/a$, $a$, $-a$, with $a=\sqrt{2}-1$. We will not review the proof here. Then we evaluate the action of the G operators as in \ref{G4}-\ref{G_6}. It might seem like we will get a multitude of terms, but in fact most of them will have more than two derivatives. The only terms that interest us are
\begin{gather}
    \left<[\tilde G_4^+V(-\frac{1}{a}),G_4^+V(\frac{1}{a})]\frac{G_4^-}{L_0}[V(a),G_4^+V(-a)]\right> = \label{massive contribution}\\ 
    2\left<[e^{-2\rho+iH}\bar d\bar DV,e^{\rho}dDV]\frac{a}{L_0}([e^{-\rho}\bar d\bar DV,e^{\rho}dDV] - i[V,\Pi^{\dot\alpha\alpha}\bar D_{\dot\alpha}D_{\alpha}V] \right.\nonumber\\
    \left.+2[V,\partial\theta^{\alpha}D_{\alpha}V])\right>
    +2\left<[\partial e^{-2\rho+iH}\bar D^2V,e^{\rho}dDV]\frac{a}{L_0}[V,\partial\theta^{\alpha}D_{\alpha}V]\right> + ... \nonumber
\end{gather}
where $\Pi^{\dot\alpha \alpha}=\partial x^{\dot\alpha \alpha}-\frac{i}{2}\left(\theta^\alpha \partial \bar{\theta}^{\dot{\alpha}}+\bar{\theta}^{\dot{\alpha}} \partial \theta^\alpha\right)$, and the ellipsis indicate higher derivative terms.

We now use the Schwinger parametrization
\begin{equation}
    \frac{1}{L_0} = \int_0^{\infty}dt\, e^{-tL_0}
\end{equation}
This will have the effect of shifting the positions of the insertions in the second commutator from $a$ and $-a$ to $b\equiv e^{-t}a$ and $-b$, plus adding an overall factor of $e^{-t}$
\begin{equation}
    \left<[A(-\frac{1}{a}),B(\frac{1}{a})]\frac{a}{L_0}[C(a),D(-a)]\right> = \int_0^{\infty}dt\, b\left<[A(-\frac{1}{a}),B(\frac{1}{a})][C(b),D(-b)]\right>
\end{equation}

We must now evaluate the correlators. Let us start with the first term on the r.h.s. of \ref{massive contribution}. Note that, since we do not want any more derivatives on the V's, we should only contract the $d$'s and $\bar d$'s among themselves. This results in, for example
\begin{gather}
    \left<\bar d_{\dot\alpha}(-\frac{1}{a})d_{\alpha}(\frac{1}{a})\bar d_{\dot\beta}(b)d_{\beta}(-b)\right> = \epsilon_{\alpha\beta}\epsilon_{\dot\alpha\dot\beta}\frac{1}{(b-\frac{1}{a})^2(b+\frac{1}{a})^2} \left(\frac{1}{2b}(-\frac{2}{a}+a(b^2+\frac{1}{a^2})) - 1 - \frac{ab}{2}\right)
\end{gather}
while the corresponding $\rho$ and $\psi$ correlators give
\begin{equation}
    \left<e^{-2\rho+iH}(-\frac{1}{a})e^{\rho}(\frac{1}{a})e^{-\rho}(b)e^{\rho}(-b)\right>_{\rho,\psi} = -\frac{8b}{a^2}\left(\frac{b-\frac{1}{a}}{b+\frac{1}{a}}\right)^3
\end{equation}
We can obtain the other orderings of the insertions by simply changing the sign of $a$ and/or $b$.

For the second term, we have
\begin{gather}
    \left<\bar d_{\dot\alpha}(-\frac{1}{a})d_{\alpha}(\frac{1}{a})\Pi^{\dot\beta\beta}(b)\right> = -i\delta_{\alpha}^{\beta}\delta_{\dot\alpha}^{\dot\beta}\left(\frac{a}{2(b-\frac{1}{a})^2} + \frac{a}{2(b+\frac{1}{a})^2} + \frac{2}{a}\frac{1}{(b^2-\frac{1}{a^2})^2}\right) \nonumber
\end{gather}

On the third term, the $d$ must contract with the $\partial\theta$, and the $\bar d$ will contract with the V's to give $\bar DV$. And on the last term, there is only one $d$, which again must contract with $\partial\theta$.

Putting everything together and integrating on $t$, we get
\begin{gather}
    \frac{1}{4}\left<[\tilde G_4^+V^0,G_4^+V^0],\frac{G_4^-}{L_0}[V^0,G_4^+V^0]\right> = \\
    -\frac{1}{4}\text{Tr}\int d^{10}xd^4\theta\, D_{\alpha}V\bar D_{\dot\alpha}VD^{\alpha}V\bar D^{\dot\alpha}V - 2D_{\alpha}V\bar D_{\dot\alpha}V\bar D^{\dot\alpha}VD^{\alpha}V \nonumber
\end{gather}
which precisely cancels the extra terms in \ref{S4 massless}. Thus the effective action from OSFT coincides with SYM to quartic order. In the next section we will argue, based on gauge invariance, that they should coincide to all orders, even though we do not explicitly compute the effective action.

\section{Gauge invariance and the full action}\label{sec Gauge invariance and the full action}
We have shown that the effective action computed from OSFT in the hybrid formalism reproduces, up to quartic order, super Yang-Mills theory plus higher derivative corrections. Differently than the Yang-Mills action, which is quartic, the SYM action \ref{MSS action} is non-polynomial in the prepotential. We would like then to generalize the result to all orders. Explicit computation of the action using the same methods as above seems impractical, as the calculations become more involved at each order. However, the SYM is actually quartic in Wess-Zumino gauge. If the gauge transformations are the same in the two theories to all orders, we can conclude that the actions must be as well.\footnote{I thank Nathan Berkovits for suggesting this idea.}

The action \ref{action} is invariant under the gauge transformations
\begin{equation}
    \delta e^V=\bar{\Sigma} e^V+e^V \Sigma, \quad \delta \Omega=-\tilde{G}_6^{+} \Sigma+[\Omega, \Sigma], \quad \delta \bar{\Omega}=-G_6^{+} \bar{\Sigma}-[\bar{\Omega}, \bar{\Sigma}]
\end{equation}
where $\bar{\Sigma}=G_4^{+} \Lambda_{-1}$ and $\Sigma=\tilde{G}_4^{+} \Lambda_2$. There are more gauge invariances, which however do not appear at the massless level \cite{Berkovits:1995ab}. Restricting the gauge parameters $\Lambda_{-1}$ and $\Lambda_2$ to be massless, i.e. $\Lambda_{-1}=e^{-\rho}\lambda(x,\theta,\bar\theta)$ and $\Lambda_{2}=e^{2\rho-iH_C}\bar\lambda(x,\theta,\bar\theta)$, we get $\bar{\Sigma}=D^2\lambda$ and $\Sigma=\bar D^2\bar\lambda$. Now, if we also truncate the string fields and their products to the massless levels (the star product simply reduces to the usual field product), these transformations reduce exactly to the SYM gauge transformations.

Now we must ask whether there are any massive contributions to the massless gauge transformations. Remember we are ignoring $\alpha'$ corrections - here this means that the string fields should appear with no derivatives in the gauge transformations, and the gauge parameters should only appear with the $D^2$ or $\bar D^2$. Note that in \ref{solution} the massless fields always appear with derivatives in the massive terms. Although we only have the explicit solution to quadratic order, we can argue that the higher orders will always include derivatives, since there is no expression we can write with only products of the string fields (with no $G$ operators, i.e. no derivatives) which is purely massive. Thus we see that the $R_i$ in \ref{decomposition} can be ignored as far as the massless gauge transformations are concerned. Massive states will still appear in the star products, but it's easy to see that also those can only contribute higher derivative terms to the gauge transformations. We therefore conclude that the effective action from OSFT has the same gauge invariance as SYM, and thus the action should be the same to all orders.

\section*{Acknowledgements}
I would like to thank Martin Schnabl for useful discussions, and for suggesting this problem. I also thank Nathan Berkovits for useful discussions, and for comments on a draft of this paper. This work was financially supported by FAPESP grant number 2023/00862-4.

\bibliographystyle{JHEP}
\bibliography{references.bib}

\end{document}